\documentclass[prb,twocolumn,floatfix,showpacs]{revtex4}

\usepackage{verbatim}   % REMOVE

\usepackage{amsmath}
\usepackage[dvips]{graphicx}
\usepackage{subfigure}
\usepackage{latexsym}
\usepackage{bm}
\usepackage{wasysym}
\usepackage{amssymb}
\usepackage{longtable}

\def\etal{~\textit{et~al.}}     % etal
\def\ra{\rangle}            % bra
\def\la{\langle}            % ket
\def\z2{$\mathbb{Z}_2$}
\def\dd{\textrm{d}}

\def\ik{{i}}
\def\jk{{j}}
\def\i{{r}}
\def\j{{r'}}
\def\rv{{\boldsymbol{r}}}
\def\rvp{{\boldsymbol{r}'}}
\def\rvo{{\hat{\boldsymbol{r}}}}
\def\Qv{{\boldsymbol{Q}}}
\def\qv{{\boldsymbol{q}}}
\def\kv{{\boldsymbol{k}}}

\newcommand{\bond}[1]{\la #1 \ra}

\newcommand{\uv}[1]{\hat{\boldsymbol{#1}}}

\newcommand{\CTRI}{
\includegraphics[width=0.3cm]{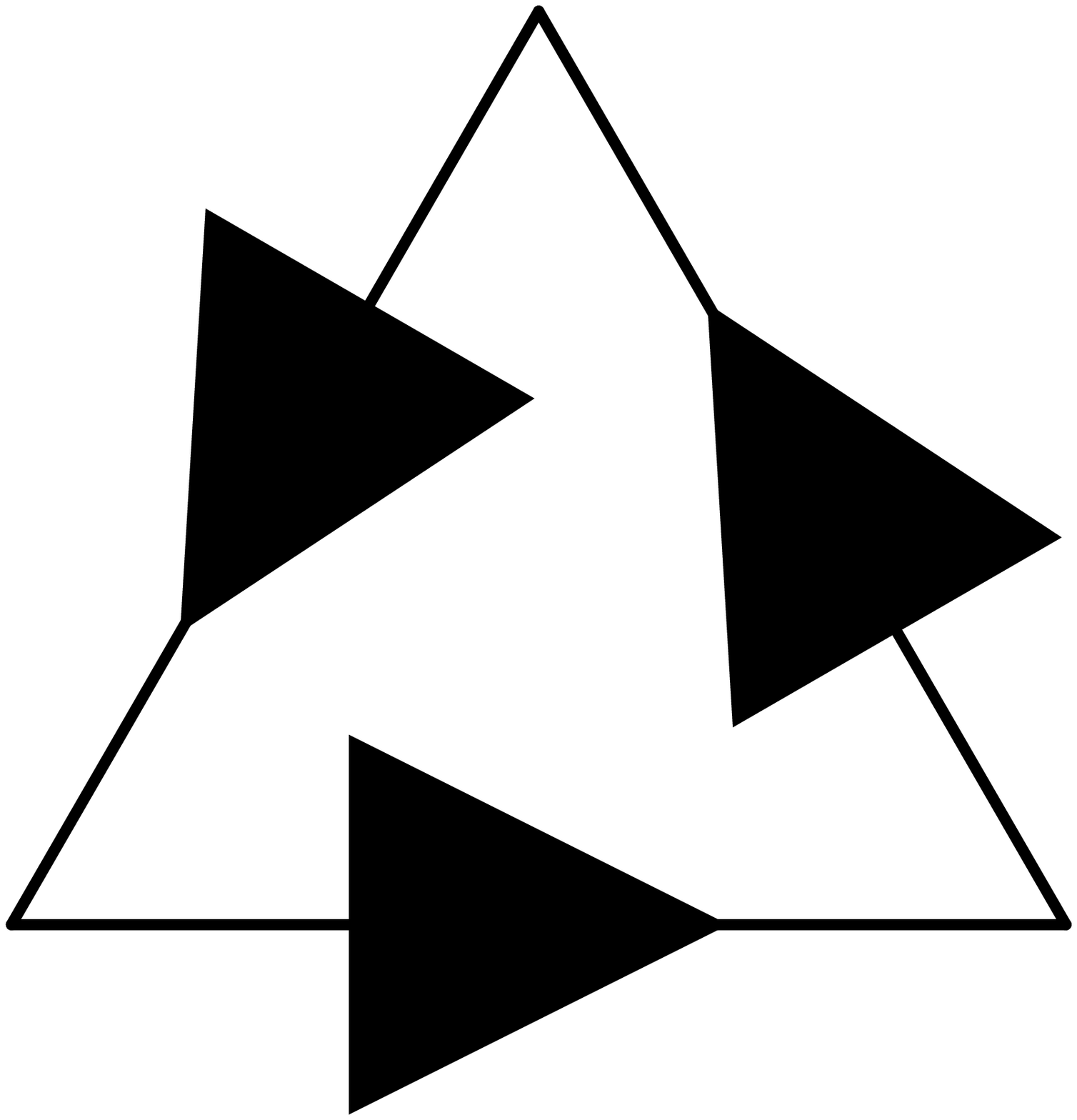}}
\newcommand{\CHEX}{
\includegraphics[width=0.4cm]{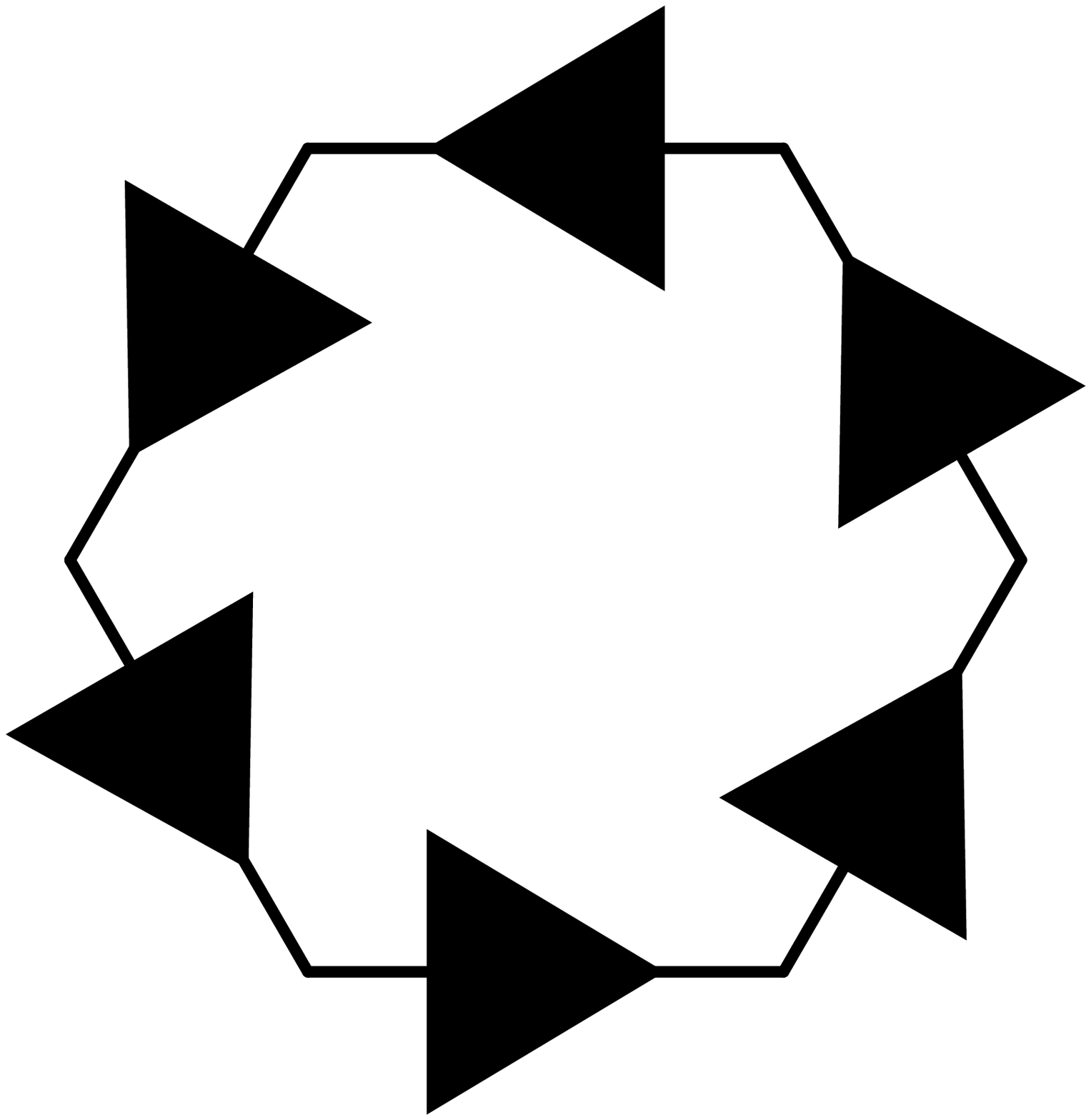}}

\newcommand{\oldnewfig}[2]{#2}

\begin{document}

\title{Theory of the Kagome Lattice Ising Antiferromagnet in Weak Transverse Field}
\author{P. Nikoli\'{c}}
\author{T. Senthil}
\affiliation{Massachusetts Institute of Technology, 77 Massachusetts Ave.,  Cambridge, MA 02139}
\date{\today}
\pacs{}

\begin{abstract}

We study the quantum Ising antiferromagnet on the Kagome lattice, with weak transverse field dynamics and other local perturbations. We analytically demonstrate the possibility of a disordered zero-temperature phase that is smoothly connected to the phase at strong transverse fields. This is done by means of an appropriate mapping to a compact U(1) gauge theory on the honeycomb lattice that is coupled to a charge-$1$ matter field. Our results are consistent with existing Monte-Carlo calculations. The differences with other commonly studied lattices (in which such disordered phases do not obtain at weak transverse fields) is explained. Ordered phases are also shown to be 
possible in principle in the weak transverse field limit, and are briefly studied.

\end{abstract}

\maketitle

\section{Introduction}

Geometrically frustrated quantum magnets hold a very important role in the active and challenging search for exotic quantum phases and spin liquids. It is hoped that the strongly competing exchange interactions between spins may in some cases prevent not only development of magnetic order, but any other long range order as well. In certain circumstances the resulting state (known as a ``spin liquid''), disordered by quantum fluctuations, could posses various unusual properties, such as non-trivial topological structure and fractionalized excitations. Much of the effort to find such fractionalized spin liquid states has been motivated by proposals of their relationship to the pseudo-gap phase of high temperature superconductors, \cite{RVB, RK, Z2} as well as many other unusual systems. \cite{Coldea, Bernu, TriRing1, TriRing2}. In addition, their topological properties seem very promising for future applications in quantum computing. \cite{QuantComp1, QuantComp2}

\begin{figure}[!b]
\includegraphics[width=2.1in,angle=90]{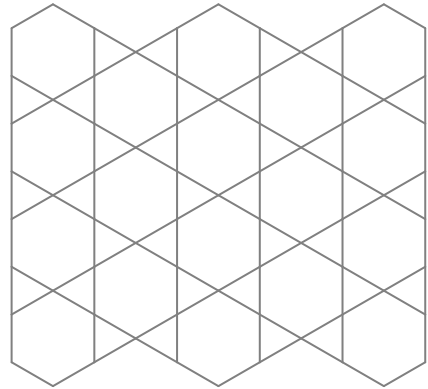}
\vskip -2mm
\caption{\label{KGrid}Kagome lattice is a corner-sharing two dimensional lattice. The frustrated units are triangular plaquettes, and they are only minimally connected into the lattice.}
\end{figure}

Out of many lattices that can host a frustrated magnet, the pyrochlore and Kagome attract most attention. Quantum liquid arising from only the nearest neighbor interactions has not yet been ruled out in these systems, unlike in many others which initially had been looked upon with great hope. The reason for this is most probably their corner-sharing structure that yields an extremely large classical degeneracy. The two-dimensional Kagome lattice (Fig.~\ref{KGrid}) has been a promising host for exotic physics for more than ten years. \cite{KagHe} Various physical states have been found at low temperatures in experiments on non-ideal Kagome magnets with spin $S \geqslant 3/2$, ranging from magnetically ordered ones in Jarosites, \cite{JarAF} to the spin-glass in $SrCr_{9p}Ga_{12-9p}O_{19}$ (SCGO) \cite{SCGOmuon, SCGOHeatCap} and newer $Ba_2Sn_2ZnGa_3Cr_7O_{22}$ (QS ferrite). \cite{QSfer} Numerical computations on small samples have provided hints of a liquid caused by quantum fluctuations in the ideal Kagome lattice spin $S=1/2$ Heisenberg model. \cite{Elser, KNum, KMagnTherm} Furthermore, in comparison to other spin systems, \cite{Lhlong, Lhshort} this one has a rather unusual spectrum, consisting of gapped spin-carrying excitations, and a band of seemingly gapless singlet states below the spin-gap. There have been several theoretical attempts to understand the unusual spectrum of the Kagome lattice quantum Heisenberg antiferromagnet, and the nature of its ground-state. Some of them favored a spin liquid, \cite{Elser, KagMila, KagSpN, KagRK, KagSpecDimer} while some other opened up a possibility of a valence bond crystal. \cite{KagSUN, KagVBC} Apparently, the Kagome lattice antiferromagnets are excellent and promising systems in which exotic phases of matter could be found.

A simpler theoretical context in which the effects of quantum dynamics on classical geometrically frustrated magnets may be studied is provided by considering frustrated Ising models perturbed by a transverse magnetic field. If the transverse field is strong (much larger than the typical exchange energy scales of the Ising interaction), disordered paramagnetic phases will result. The interesting and non-trivial questions therefore arise in the limit of weak transverse fields. What is the effect of such a weak transverse field on the classical Ising magnet? Generally a number of different outcomes are possible. Consider the rather common situation where the classical Ising magnet has a macroscopic number of degenerate ground states, and no long range magnetic order even at zero temperature. In some such cases a weak transverse field is known to immediately select a particular ordered ground state from this degenerate manifold. This phenomenon, known as ``order-by-disorder'', happens for instance for the triangular Ising antiferromagnet, or the fully frustrated square lattice Ising model in weak transverse fields. In contrast, on the Kagome lattice Monte-Carlo calculations \cite{FrIsing1,FrIsing2} report that the classical Ising antiferromagnet stays disordered even in the presence of weak transverse fields. This phenomenon has been dubbed ``disorder-by-disorder''.

Our interest in this paper is in elucidating better the physics of the Kagome Ising antiferromagnet in weak transverse fields, and possibly other local perturbations consistent with the Ising symmetry. Our particular focus will be on obtaining some analytical understanding of the disordered paramagnetic phase found in numerical calculations. To that end we develop a reformulation of the model in terms of an appropriate gauge theory that lives on the honeycomb lattice. General structure of the phases of this gauge theory will be analyzed through duality transformations. Our approach enables us to understand differences between the Kagome lattice and the analogous problems on the square or triangular lattices where order-by-disorder seems to occur.

Besides its intrinsic interest, the problem studied in this paper also serves as a useful technical playground for learning about some of the fundamental issues that arise in analytical studies of frustrated magnets. Theoretical demonstrations of quantum liquid-like behavior in specific microscopic models of spin or boson systems are rare, and currently involve interactions beyond the nearest neighbor. \cite{KagFract, BosonFract1, BosonFract2, BosonFract3,PyroU1}. The ideas developed in this paper may perhaps motivate construction of simpler microscopic models that display exotic phenomena such as fractionalization. We emphasize however that the paramagnetic ``liquid'' phase discussed in the present paper is non-exotic, and is expected to be smoothly connected to that which obtains in the large transverse field limit.

\section{Lattice Field Theory of Fluctuating Spins}

The basic model that we want to study is the antiferromagnetic Ising model in transverse magnetic field on the Kagome lattice (TFIM):
\begin{equation}\label{TFIM}
H = J \sum_{\bond{\ik\jk}} S^z_\ik S^z_\jk - \Gamma \sum_\ik S^x_\ik \ .
\end{equation}
Here $\ik, \jk$ label the sites of the Kagome lattice, and $\boldsymbol{S}_\ik$ is a spin $S=\frac{1}{2}$ moment at site $\ik$. $J$ is the antiferromagnetic Ising interaction strength, and $\Gamma$ is the strength of the transverse magnetic field. More generally, we will focus on the regime where the energy scale $J$, which fixes an easy axis, is much larger than all other energy scales ($J \gg \Gamma$), and consider various ways in which dynamics can be given to the spins, without conserving any quantities (clearly, the alternate limits of large $\Gamma$ is trivial). The main goal is to study the structure of possible phases that can emerge when the frustrated Ising antiferromagnet is endowed with weak quantum dynamics (that preserves Ising, but no other spin symmetries).

Our strategy is as follows. We derive a low energy effective theory that is appropriate in the easy axis limit. This may be represented as a compact U(1) gauge theory on the honeycomb lattice that is coupled to a bosonic ``matter'' field (with gauge charge $1$). There is in addition a non-zero static \emph{background} charge at each site. The utility of a compact U(1) gauge theory (with appropriate background charges) to describing the low energy physics of frustrated easy-axis magnets has been pointed out
several times in the literature. However, in contrast to the Kagome lattice, on other lattices typically the gauge theory has no dynamical matter fields. In two spatial dimensions on these other lattices the gauge theory is in a confined phase, and the presence of background charges leads to broken translation symmetry. Presence of the additional dynamic matter field distinguishes the Kagome lattice from these other lattices. As we will see, it is now possible to have a phase that is also ``confining'' (more precisely a Higgs phase), which preserves translation symmetry even in the presence of the background charges. For the original Kagome TFIM this describes a translation invariant paramagnet. The general possibility of such translation invariant Higgs phases in such gauge theories has been discussed before.\cite{SJ}

We analyze the gauge theory appropriate for the Kagome TFIM using duality transformations. The dual theory will turn out to be equivalent to a certain XY model with a three-fold anisotropy. The disordered phase of this model is the Higgs phase discussed above. The ordered phase describes a situation where there is order-by-disorder in the original Kagome magnet.

\subsection{U(1) Gauge Theory}

\begin{figure}
\includegraphics[width=2.6in]{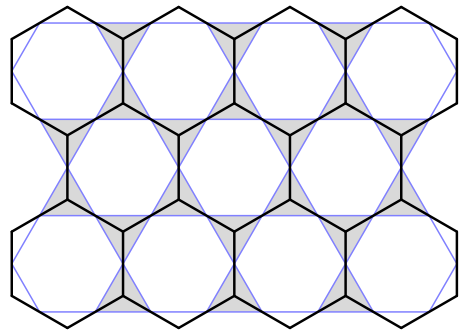}
\vskip -2mm
\caption{\label{KagHon}Relationship between the honeycomb and Kagome lattices.
Every honeycomb bond contains one Kagome site, while every Kagome triangle contains a honeycomb site.}
\end{figure}

The nearest neighbor Ising coupling on the Kagome lattice can be conveniently written as a sum of terms defined on the triangular plaquettes:
\begin{equation}\label{KagIsing}
H_z = J \sum_{\bond{\ik\jk}} S^z_\ik S^z_\jk =
  \frac{J}{2} \sum_{\triangle} \Bigl( \sum_{\ik \in \triangle} S^z_\ik \Bigr)^2
  + \textrm{const.} \ .
\end{equation}
This allows us to easily describe the sector of low energy states: total spin on every triangle should be $\pm 1/2$. Spin configurations that satisfy this condition are \emph{least frustrated}. Let us express these states using a set of variables defined on the \emph{honeycomb} lattice, whose sites we will label by $p$ and $q$. Figure \ref{KagHon} illustrates relationship between the Kagome and honeycomb lattices. The honeycomb bonds contain Kagome sites, and we can associate with them the Kagome spins: $S^z_\ik \equiv S^z_{\bond{pq}}$. On the other hand, the honeycomb sites reside inside the Kagome triangular plaquettes. It is useful to keep track of the total spin (whether it is $+1/2$ or $-1/2$) on any triangular Kagome plaquette. We therefore introduce a variable $s^z_p$ that measures the total plaquette spin:
\begin{equation}\label{TriSpin}
(\forall p) \quad s^z_p = \sum_{q \in p} S^z_{\bond{pq}} \ .
\end{equation}
We impose restriction to the low energy sector by requiring that $s^z_p$ take only values $\pm \frac{1}{2}$. Note that when a Kagome spin is flipped (provided that flipping does not introduce more frustration), $s^z_p$ on the triangles that contain it change sign
in the same direction. With only the Ising interaction present there is a large number of classical ground states: every spin configuration that satisfies Equation ~(\ref{TriSpin}) with $s^z_p = \pm 1/2$ is a classical ground state. Inclusion of the $\Gamma$ or other terms in the Hamiltonian will split this huge degeneracy of the ground
state manifold.

It is possible to perturbatively construct an effective theory that describes the low energy dynamics in the ground state manifold, and express it on the honeycomb lattice:
\begin{eqnarray}\label{HonEff}
H_{\textrm{eff}} & = & -\frac{\Gamma}{2} \sum_{\bond{pq}}
         \Bigl(s^+_p S^+_{\bond{pq}} s^+_q + h.c. \Bigr) - \cdots \\
   & - & t \sum_{\hexagon}
         \Bigl( S^+_{\bond{12}} S^-_{\bond{23}} S^+_{\bond{34}}
                S^-_{\bond{45}} S^+_{\bond{56}} S^-_{\bond{61}} \nonumber \\
   & + & h.c. \Bigr) - \cdots \nonumber \ .
\end{eqnarray}
The Hilbert space of this theory is defined only by the least frustrated states. The lowest order dynamical term is a single Kagome spin flip $S^{\pm}$ caused by the transverse field, and the operators $s^{\pm}$ simply project-out the states that are not minimally frustrated. At higher orders of perturbation theory multiple spins are being flipped, and the smallest ``ring-exchange'' term appears at the sixth order, so that $t \sim \frac{\Gamma^6}{J^5}$. Alternately, we could have imagined adding such a ring-exchange term to the original model (in addition to the transverse field term). We will examine the properties of this effective Hamiltonian in the low energy subspace for arbitrary $\Gamma$ and $t$. The pure transverse field model then corresponds to the particular limit $\Gamma \gg t$. All fluctuations are constrained by ~(\ref{TriSpin}). The effective theory will have this general form for all kinds of Kagome spin models with an easy axis, provided that dynamics preserves only the Ising symmetry.

\begin{figure}
\oldnewfig
{\includegraphics[width=2.6in]{honeycomb-charge.eps}}
{\includegraphics[width=1.8in, viewport=80 310 240 520, angle=-90, clip]{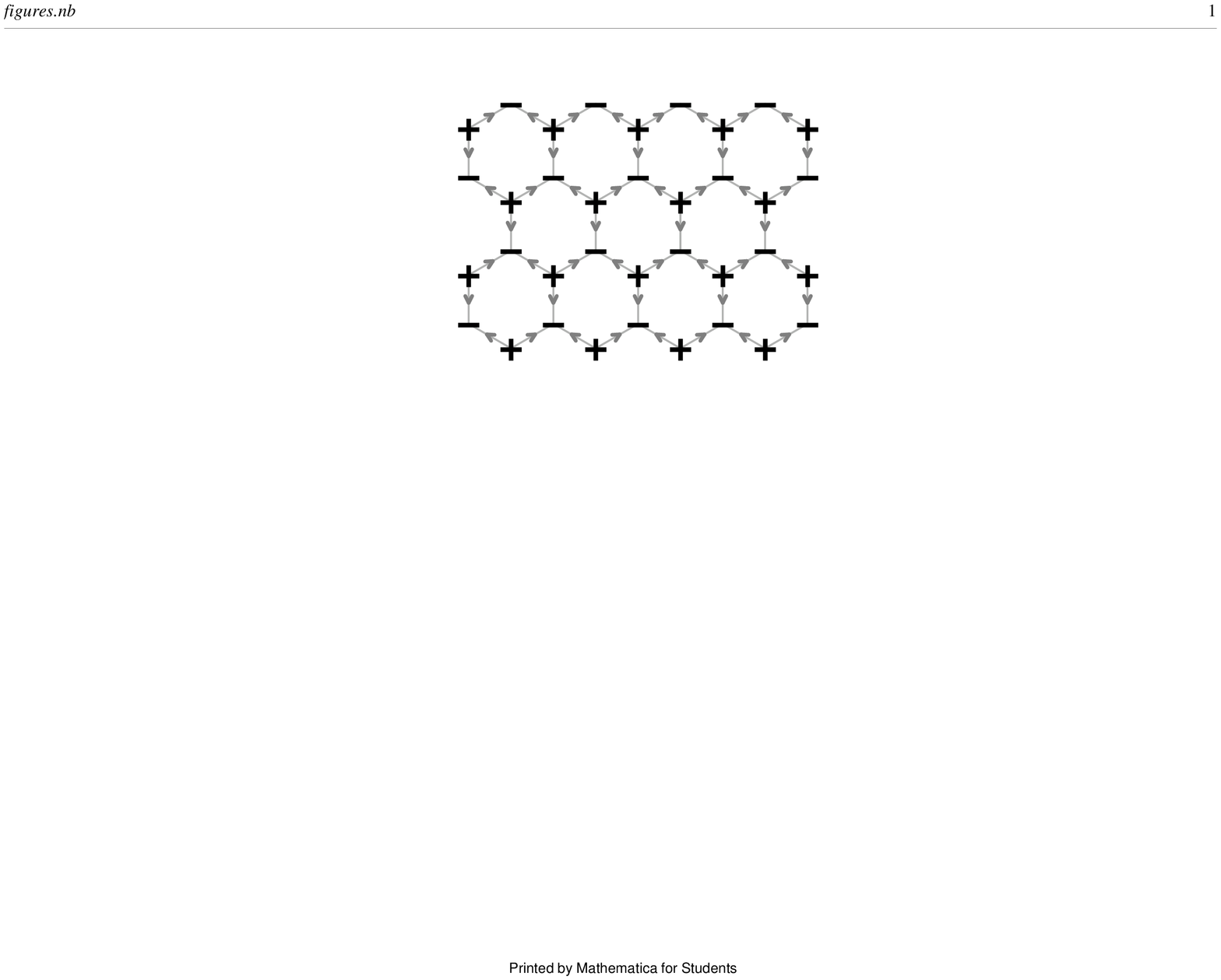}}
\caption{\label{HonCharge}Fixed background charge $\varepsilon_p = \pm 1$ on the honeycomb sites, and the electric field $E^{(0)}_{pq}$ created by it (up to a multiplicative factor). On every honeycomb bond we define $E^{(0)}_{pq} = \varepsilon_p/2 = -E^{(0)}_{qp} = -\varepsilon_q/2$ ($\textrm{div}E^{(0)} = \frac{3}{2}\varepsilon_p$).}
\end{figure}

The Hamiltonian ~(\ref{HonEff}) can be interpreted as a model of charged bosons moving in presence of a fluctuating electromagnetic field. In order to formulate this interpretation, we need to exploit the bipartite nature of the honeycomb lattice. Let us introduce a fixed field $\varepsilon_p$ that takes values $+1$ and $-1$ on two different sublattices, as shown in Fig.~\ref{HonCharge}. Then, define:
\begin{eqnarray}\label{EnDef}
n_p & = & \varepsilon_p \Bigl( s^z_p + \frac{1}{2} \Bigr) \\
E_{pq} & = & \varepsilon_p \Bigl( S^z_{\bond{pq}} + \frac{1}{2} \Bigr) \nonumber \ .
\end{eqnarray}
Note that $E_{pq} = - E_{qp}$, so that it can be viewed as a vector living on the bonds of the honeycomb lattice. This will be interpreted as an integer-valued ``electric'' field. The integer $n_p$ will be interpreted as the gauge charge of a bosonic ``matter'' field that couples to the electric field. The constraint ~(\ref{TriSpin}) becomes:
\begin{equation}\label{HonGauss}
(\forall p) \quad \sum_{q \in p} E_{pq} = n_p + \varepsilon_p \ ,
\end{equation}
where the sum on the left-hand side is taken over three honeycomb sites neighboring to $p$. Natural interpretation of this equation is Gauss' Law: divergence of the electric field $E_{pq}$ is equal to the total local charge. Note that the boson occupation numbers $n_p$ take values $0$ and $\varepsilon_p$. There is an additional fixed background charge distribution $\varepsilon_p$.

At this level, the boson occupation number and electric field strength are constrained to only two integer values by ~(\ref{EnDef}). It is useful to soften this ``hard-core'' by allowing $n_p$ and $E_{pq}$ to take arbitrary integer values, but penalizing fluctuations where either quantity assumes values different from that dictated by the hard-core condition. It will also be useful to introduce the corresponding conjugate operators, $\varphi_p$ and $\mathcal{A}_{pq}$, which are angular variables in $[0,2\pi)$. The boson creation and annihilation operators $s^{\pm}_p$ simply become $\exp(\pm i\varepsilon_p\varphi_p)$, and similar holds for the electric field: $S^{\pm}_{\bond{pq}} \rightarrow \exp(\pm i\varepsilon_p\mathcal{A}_{pq})$. Now it is straight forward to rewrite the Hamiltonian ~(\ref{HonEff}) as a compact U(1) gauge
theory (up to a constant):
\begin{eqnarray}\label{HonU1}
H & = & U_1 \sum_{\bond{pq}} \Bigl( E_{pq} - E^{(0)}_{pq} \Bigr)^2 +
        U_2 \sum_p \Bigl( n_p - \frac{\varepsilon_p}{2} \Bigr)^2 \nonumber \\
  & - & \Gamma\sum_{\bond{pq}} \cos \Bigl( \varphi_q - \varphi_p - \mathcal{A}_{pq} \Bigr) - \cdots \\
  & - & t \sum_{\hexagon} \cos \Biggl( \sum_{\bond{pq}}^{\CHEX_{}} \mathcal{A}_{pq} \Biggr)
        - \cdots \nonumber \ .
\end{eqnarray}
We have labeled by $E^{(0)}_{pq}$ a fixed background electric field that originates due to the background charge (see Fig.\ref{HonCharge}). The terms proportional to $U_1$ and $U_2$ penalize fluctuations of the boson number and the electric field away from the preferred ``hard-core'' values.

We see that $\mathcal{A}_{pq}$ plays role of a vector potential. The term at the lowest order of perturbation theory describes boson hopping on the honeycomb lattice, while the term at the sixth order gives energy cost to the ``magnetic flux'' (away from $2\pi \times \textrm{integer}$ values). It is also easy to write down other terms consistent with the symmetries and the $U(1)$ gauge structure. Fluctuations are subject to the Gauss' Law ~(\ref{HonGauss}). Note that the charged bosons cannot screen out the background charge ($n_p \to \lbrace 0,\varepsilon_p \rbrace$) without paying a large price ($U_1$,$U_2$). This is a consequence of magnetic frustration in the original Kagome spin model. In fact, there is a large number of degenerate boson configurations that accomplish the best possible screening, and they correspond to the least frustrated states. However, the very fact that there is a matter field in this compact U(1) gauge theory distinguishes the Kagome lattice from other commonly studied lattices. It gives hope that a translation symmetric paramagnetic phase could exist on the Kagome lattice, in contrast to the situation for the triangular Ising antiferromagnet, or the fully frustrated square lattice Ising magnets (with weak transverse field dynamics).

\subsection{Duality transformation}

\begin{figure}
\includegraphics[height=2.0in]{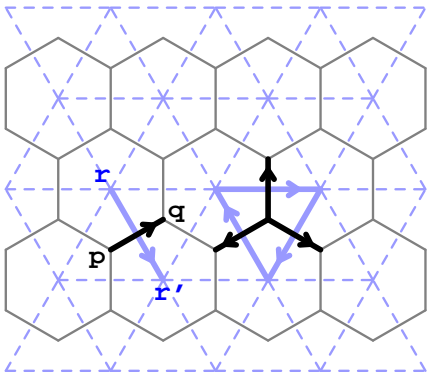}
\caption{\label{HonTriDuality}Duality between the honeycomb and triangular lattices. The triangular lattice sites sit inside the honeycomb plaquettes, and vice versa. Duality between a honeycomb lattice vector $p \to q$ and a triangular lattice vector $\i \to \j$ is shown in the lower-left portion of the graph: their directions are related by the right-hand rule. Divergence on the honeycomb lattice translates into negative lattice curl on the triangular lattice (we always take circulation in the counter-clockwise sense). }
\end{figure}

In this section we perform a duality transformation on the compact $U(1)$ gauge theory derived above. This will enable us to analyze the structure of the possible phases. The duality transformation proceeds in a standard fashion. We first derive the path-integral form of the compact U(1) gauge theory. All terms denoted by ellipses in ~(\ref{HonU1}) will be ignored. The action will contain a usual Berry's phase (we will omit the time index):
\begin{equation}\label{HonBerry}
S_{\textrm{B}} = -i \sum_{\tau} \Biggl( \sum_{\bond{pq}} \mathcal{A}_{pq} 
  \Delta_{\tau} E_{pq} + \sum_{p} \varphi_{p} \Delta_{\tau} n_{p} \Biggr) \ ,
\end{equation}
and a potential energy part:
\begin{equation}\label{HonPot}
S_{\textrm{P}} = \sum_{\tau}\delta\tau \Biggl[
  U_1 \sum_{\bond{pq}}\Bigl( E_{pq}-E^{(0)}_{pq} \Bigr)^2 +
  U_2 \sum_p \Bigl( n_p - \frac{\varepsilon_p}{2} \Bigr)^2 \Biggr] \nonumber
\end{equation}
where $\delta\tau$ is the imaginary time increment. The kinetic energy part will be obtained by applying Villain's approximation to the cosine terms in ~(\ref{HonU1}):
\begin{equation}\label{Villain}
e^{t \cos\theta} \approx \sum_{m=-\infty}^{\infty} e^{-K m^2 - i m \theta}
  \quad , \quad t = 2e^{-K} \to 0 \ .
\end{equation}
Two new integer-valued Villain fields will appear: a particle current $j_{pq}$, and a magnetic field scalar $B_\i$ that lives inside plaquettes of the honeycomb lattice, or equivalently on the dual triangular lattice sites (see Fig.~\ref{HonTriDuality}):
\begin{eqnarray}\label{HonKin}
S_{\textrm{K}} & = & \sum_{\tau} \Biggl[
        K_1 \sum_{\hexagon_\i} B_\i^2 + K_2 \sum_{\bond{pq}} j_{pq}^2 \\
  & + & i\sum_{\bond{pq}} j_{pq} \Bigl( \varphi_q - \varphi_p - \mathcal{A}_{pq} \Bigr)
      + i\sum_{\hexagon_\i} B_\i \Biggl( \sum_{\bond{pq}}^{\CHEX_\i} \mathcal{A}_{pq} \Biggr)
  \Biggr] \nonumber \ .
\end{eqnarray} 
We will treat the constants $K_1$ and $K_2$ as free parameters (they are in principle determined in terms of the microscopic parameters that define the original Kagome Hamiltonian). Our interest is in exploring the general nature of the possible phases that are contained in this dual action. The angular variables $\varphi_p$ and $\mathcal{A}_{pq}$ can now be formally integrated out, yielding Kronecker-delta factors in the path-integral for all integer-valued expressions that they couple to. Such factors simply express familiar laws of electrodynamics. The integral over the boson phase angle will give rise to the current conservation law:
\begin{equation}\label{HonCurrCon}
\int_0^{2\pi}\dd\varphi_p \qquad \longrightarrow \qquad
  \Delta_{\tau} n_{p} + \sum_{q \in p} j_{pq} = 0 \ ,
\end{equation}
while the integral over the vector potential will reproduce a two-dimensional Maxwell's equation:
\begin{equation}\label{HonMaxwell}
\int_0^{2\pi}\dd\mathcal{A}_{pq} \qquad \longrightarrow \qquad
  \Delta_{\tau}E_{pq} + j_{pq} = B_\i - B_\j \ .
\end{equation}
The direction of the triangular lattice vector $B_\i - B_\j$ in the last equation is related to the direction of $E_{pq}$ by the right-hand rule (see Fig.~\ref{HonTriDuality}).

The equations ~(\ref{HonGauss}), ~(\ref{HonCurrCon}), and ~(\ref{HonMaxwell}) can be solved on the dual triangular lattice, whose sites will be labeled by $\i$ and $\j$. As is standard, we first define spatial components of a dual gauge field $A_{\i\j}$ on the dual triangular lattice, and a fixed background gauge field $A^{(0)}_{\i\j}$, such that their circulations on a triangular plaquette are determined by the charge contained ``inside'' the plaquette:
\begin{equation}\label{TriVectSpace}
(\forall p) \quad \sum^{\CTRI_p}_{\bond{\i\j}} A_{\i\j} = n_p \quad , \quad
  \sum^{\CTRI_p}_{\bond{\i\j}} A^{(0)}_{\i\j} = \varepsilon_p \ .
\end{equation}
The background gauge field $A^{(0)}_{\i\j}$ can be specified in many different ways, and we will discuss a convenient choice later. For now, we will only assume that it does not depend on imaginary time. If we substitute these definitions into the Gauss' Law ~(\ref{HonGauss}), and note that the honeycomb lattice divergence of the electric field translates into the negative triangular lattice curl, we can solve it by writing:
\begin{equation}\label{HonGaussSol}
E_{pq} = \chi_\i - \chi_\j - A_{\i\j} - A^{(0)}_{\i\j} \ .
\end{equation}
Again, orientation of the vectors on the right-hand side is related to that of $E_{pq}$ by the right-hand rule. The new field $\chi_\i$ is an allowed degree of freedom, since lattice curl of a pure gradient vector field is zero. Care must be taken to ensure the integer-valued nature of $E_{pq}$. A convenient way to enforce this will be discussed below. Let us also define temporal components of the dual gauge fields:
\begin{equation}\label{TriVectTime}
(\forall \i) \quad A_{\i,\i+\hat{\tau}} = B_\i - \Delta_{\tau}\chi_\i \quad , \quad
   A^{(0)}_{\i,\i+\hat{\tau}} = 0 \ .
\end{equation}
Substituting this and ~(\ref{HonGaussSol}) into ~(\ref{HonMaxwell}) gives us an expression for the particle current:
\begin{equation}\label{HonCurrSol}
j_{pq} = A_{\i,\i+\hat{\tau}} - A_{\j,\j+\hat{\tau}} + \Delta_{\tau}A_{\i\j} =
  \hat{r}_{pq} \cdot \textrm{curl}A \ .
\end{equation}
Therefore, the current becomes expressed as lattice curl of the dual vector potential, taken on the triangular lattice temporal plaquette that is pierced by the current vector $\boldsymbol{j} = j \hat{\rv}$. Finally, we note that the current conservation ~(\ref{HonCurrCon}) is not independent from the Gauss' Law ~(\ref{HonGauss}) and the Maxwell's equation ~(\ref{HonMaxwell}).

\begin{figure}
\includegraphics[height=2.0in]{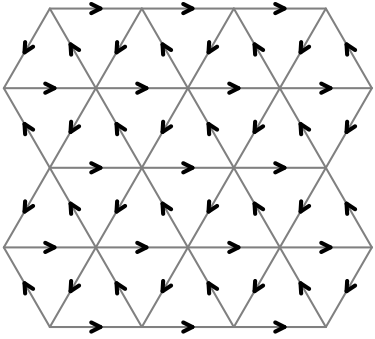}
\caption{\label{TriVectBackground}Convenient definition of the fixed background gauge field $A^{(0)}_{\i\j}$. On every bond $A^{(0)}_{\i\j} = 1/3$ in the given direction $\i \to \j$, so that its (counter-clockwise) circulations yield $\varepsilon_p$ on every plaquette. This vector field is formally dual to that of $E^{(0)}_{pq}$ on the honeycomb lattice (up to a factor).}
\end{figure}

\begin{figure}
\subfigure[{}]{\includegraphics[width=1.4in]{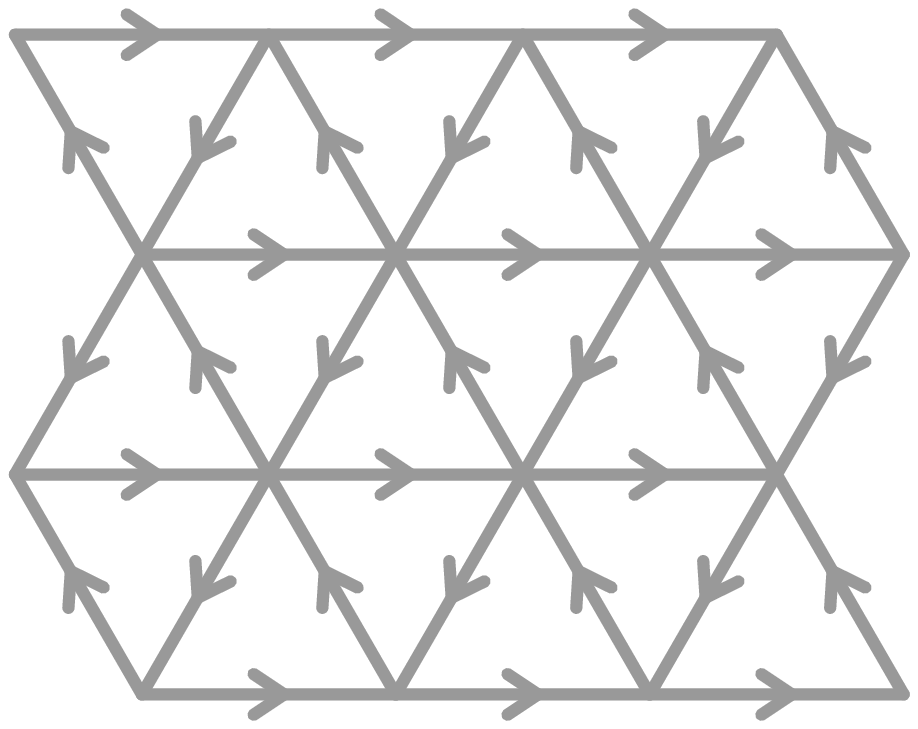}}
\subfigure[{}]{\includegraphics[width=1.4in]{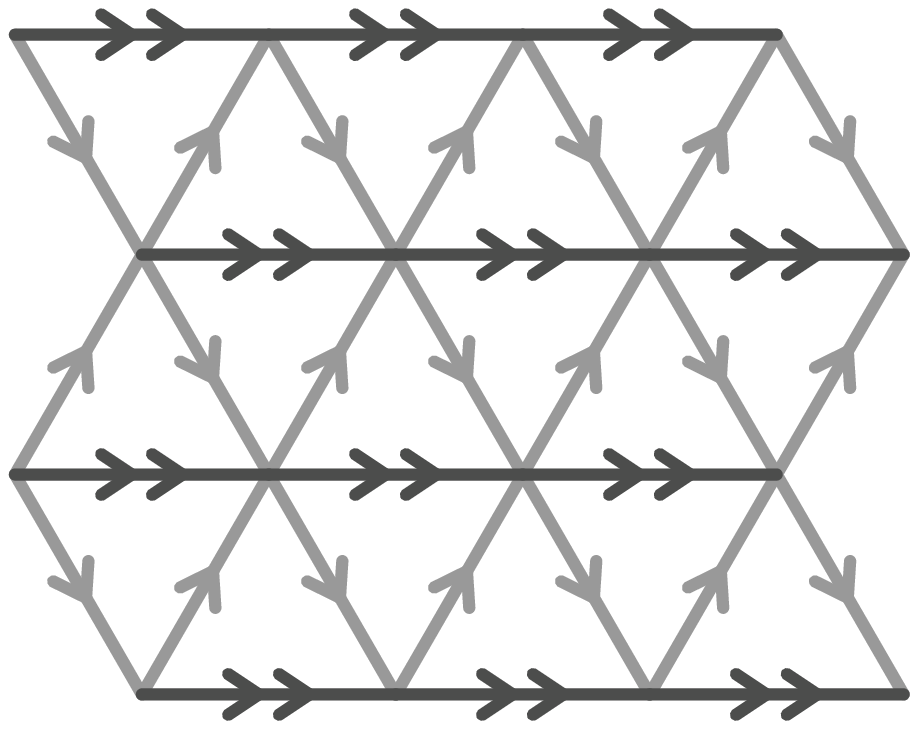}}
\vskip -2mm
\subfigure[{}]{\oldnewfig
{\includegraphics[width=1.2in]{tri-A0-4.eps}}
{\includegraphics[width=1.2in, angle=-90, viewport=70 310 250 530, clip]{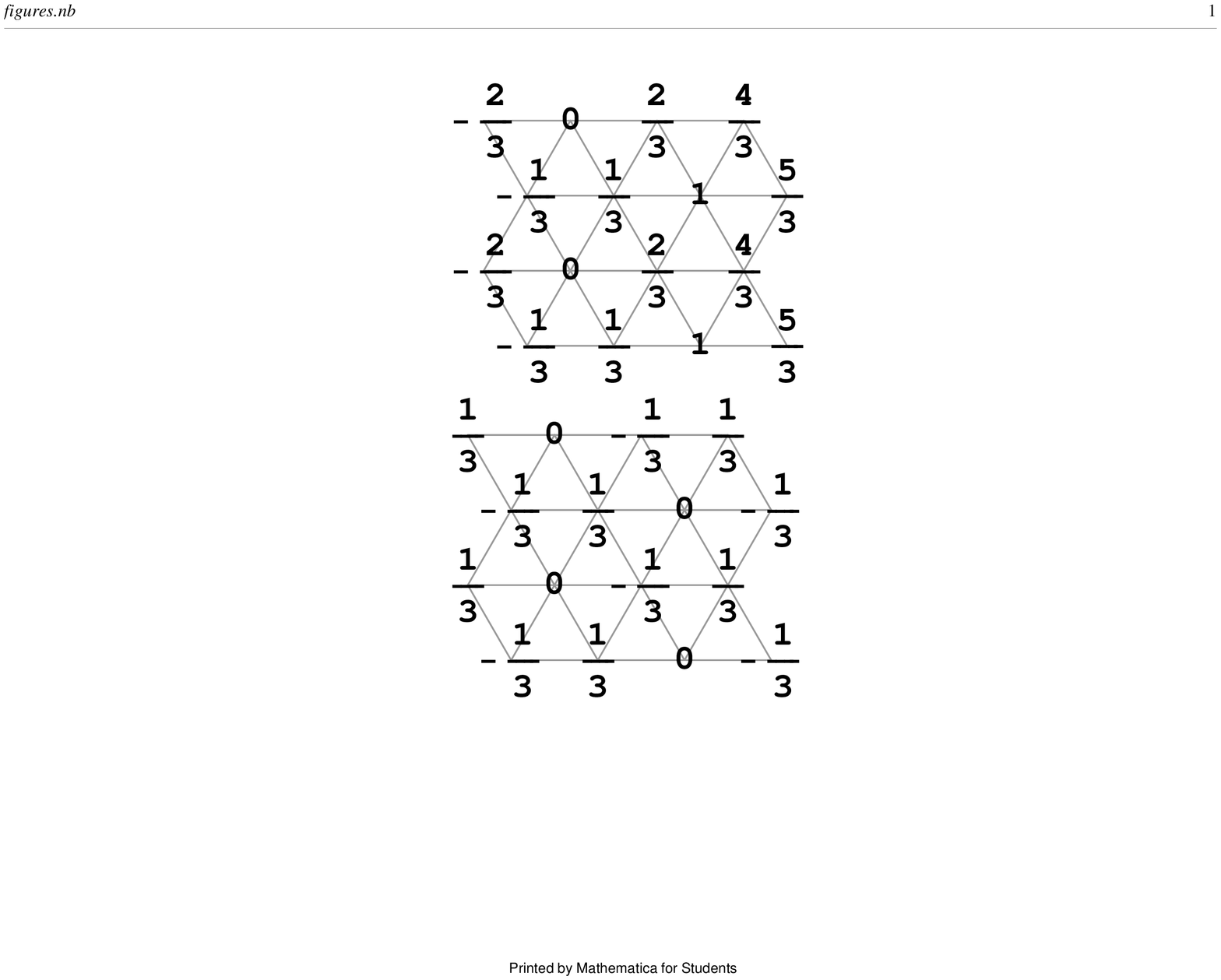}}}
\subfigure[{}]{\oldnewfig
{\includegraphics[width=1.2in]{tri-A0-5.eps}}
{\includegraphics[width=1.2in, angle=-90, viewport=248 310 428 530, clip]{specfig2.eps}}}
\vskip -2mm
\caption{\label{TriGaugeDecomp}(a) Fixed background gauge field $A^{(0)}_{\i\j}$. (b) A curl-less vector field $A'_{\i\j}$. Every arrow contributes $1/3$ in the given direction, so that the sum $A^{(0)}_{\i\j} + A'_{\i\j}$ is an integer-valued vector field. (c) A scalar field $\chi'_\i$ whose gradient gives the curl-less vector field: $A'_{\i\j} = \chi'_\j - \chi'_\i$. (d) Fractional parts $\chi^{(0)}_\i$ of the scalar field $\chi'_\i$. Requiring that the fields $\chi_\i$ in ~(\ref{HonGaussSol}) have fractional parts equal to $\chi^{(0)}_\i$ compensates for fractional values of $A^{(0)}_{\i\j}$, and makes $E_{pq}$ an integer.}
\end{figure}

We can now obtain dual form of the following action that remained after integrating out the conjugate angles:
\begin{eqnarray}
S & = & \sum_{\tau} \Biggl[
        U_1\delta\tau \sum_{\bond{pq}} \Bigl( E_{pq}-E^{(0)}_{pq} \Bigr)^2 +
        K_1 \sum_{\hexagon_\i} B_\i^2 \nonumber  \\
  & & + U_2\delta\tau \sum_p \Bigl( n_p - \frac{\varepsilon_p}{2} \Bigr)^2 +
        K_2 \sum_{\bond{pq}} j_{pq}^2 \Biggr]
\end{eqnarray}
We will eliminate $E_{pq}$, $B_\i$, $n_p$, and $j_{pq}$ using ~(\ref{HonGaussSol}), ~(\ref{TriVectTime}), ~(\ref{TriVectSpace}), and ~(\ref{HonCurrSol}) respectively. The particle number and current terms together yield curls of the dual vector potential on spatial and temporal plaquettes of the triangular lattice. Though not necessary, for simplicity we set $U_2\delta\tau = K_2 = g/2$, and $U_1\delta\tau = K_1 = e^2/2$ and consider the phase diagram in the resulting section of coupling constant space. In order to complete the duality transformation, we must also translate the quantities $E^{(0)}_{pq}$, and $\varepsilon_p$ into the dual language. A natural dual counterpart of the background electric field vector $E^{(0)}_{pq}$ is $A^{(0)}_{\i\j}$, since both are determined by the background charge distribution $\varepsilon_p$. In the spirit of equation ~(\ref{HonGaussSol}) this suggests the identification $E^{(0)}_{pq} = -\frac{3}{2}A^{(0)}_{\i\j}$, shown in the Fig.~\ref{TriVectBackground}. However, the values of $A^{(0)}_{\i\j}$ are then not integers, and compensation is necessary in order to keep $E_{pq}$ integer-valued in ~(\ref{HonGaussSol}). A convenient solution is to require that $\chi_\i$ fields take particular non-integer and site-dependent set of values. Specifically, we demand that $\chi_\i - \chi^{(0)}_\i$ be integers, with the fixed fractional offsets $\chi^{(0)}_\i$ as illustrated in the Fig.~\ref{TriGaugeDecomp}. Then, the dual action is:
\begin{eqnarray}\label{TriAction}
S & = & \frac{e^2}{2}\sum_{\bond{\i\j}}
        \Bigl(\chi_\i-\chi_\j-A_{\i\j}+\frac{1}{2}A^{(0)}_{\i\j}\Bigr)^2 \nonumber \\
  & & + \frac{g}{2}\sum_{\textrm{plaq.}}
        \Bigl(\textrm{curl}\Bigl(A-\frac{1}{2}A^{(0)}\Bigr)\Bigr)^2 \ ,
\end{eqnarray}
where the first summation extends over all space-time links, and the second over all spatial and temporal plaquettes. Let us also shift
\begin{equation}
A \rightarrow A - \frac{1}{2}A_0 \ .
\end{equation}
The action then reads:
\begin{equation}\label{TriAction2}
S = \frac{e^2}{2} \sum_{\bond{\i\j}} \Bigl( \chi_\i - \chi_\j - A_{\i\j} \Bigr)^2
  + \frac{g}{2}\sum_{\textrm{plaq.}}\bigl(\textrm{curl}A\bigr)^2 \ .
\end{equation}
In this theory $\chi_\i - \chi^{(0)}_\i$, and $A_{\i\j} + A^{(0)}_{\i\j}/2$ take integer values. The fractional residua $\chi^{(0)}_\i$ are given in the Fig.~\ref{TriGaugeDecomp}(d).

\begin{figure}
\subfigure[{}]{\includegraphics[height=1.2in]{triangular-staggered-flux.eps}}
\subfigure[{}]{\includegraphics[height=1.2in]{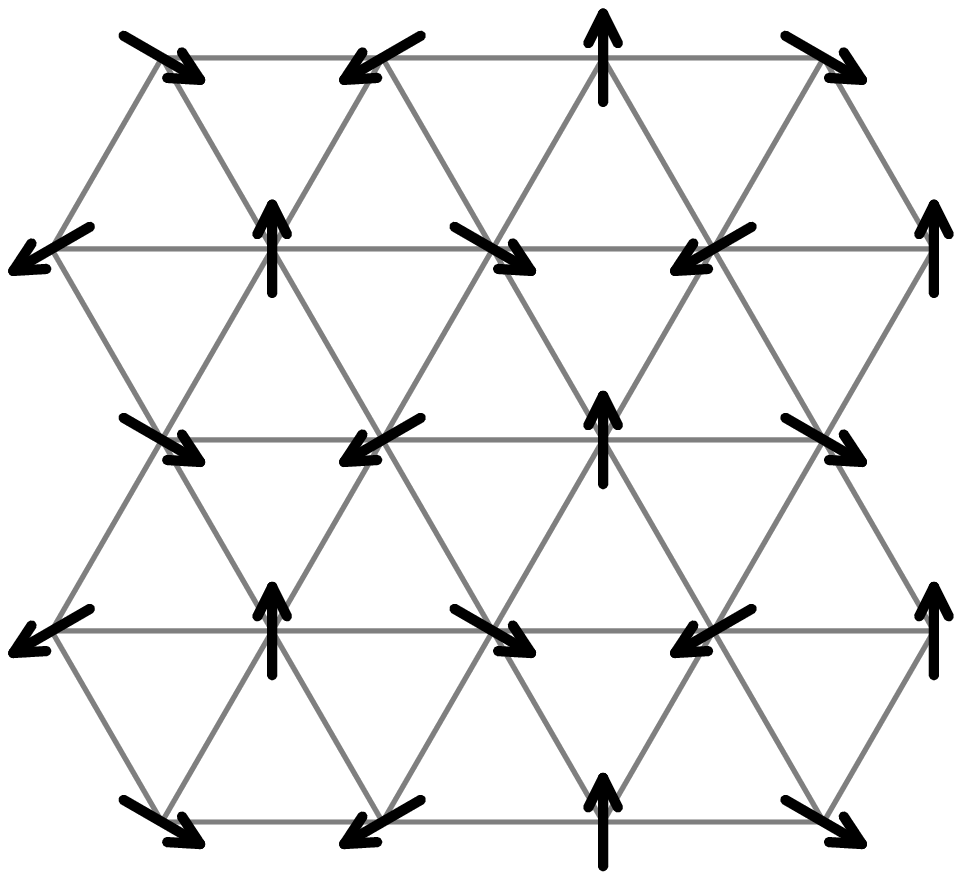}}
\caption{\label{XYExtField}(a) Field lines of the background gauge field $A^{(0)}_{\i\j}/2$ that produces a staggered flux. (b) External XY field $2\pi\chi^{(0)}_\i$ that breaks XY and translational symmetries (compare with Fig.~\ref{TriGaugeDecomp}(d)). This field is constant on any given sublattice of the triangular lattice.}
\end{figure}

It will actually be convenient to impose these integer conditions on the $\chi_\i$ and $A_{\i\j}$ fields softly (in the same spirit as the usual sine-Gordon description of Coulomb gases). The result is a generalized sine-Gordon theory with the structure:
\begin{eqnarray}\label{TriSineGordon}
S & = & \frac{e^2}{2} \sum_{\bond{\i\j}} \Bigl(\chi_\i - \chi_\j - A_{\i\j} \Bigr)^2 
      + \frac{g}{2}\sum_{\textrm{plaq.}}\bigl(\textrm{curl}A\bigr)^2 \nonumber \\
  & & - K \sum_{\bond{\i\j}} 
        \cos 2\pi\Bigl( A_{\i\j} + \frac{1}{2} A^{(0)}_{\i\j} \Bigr) - \cdots \nonumber \\
  & & - \gamma \sum_\i \cos 2\pi\bigl( \chi_\i - \chi^{(0)}_\i \bigr) - \cdots \ .
\end{eqnarray}
In principle, higher harmonics of the lowest order cosine terms shown above must also be included. The fields $\chi_\i$ and $A_{\i\j}$ now take real values. Let us absorb the field $\chi_\i$ into $A_{\i\j}$ by the shift:
\begin{equation}
A_{\i\j} \rightarrow A_{\i\j} + \chi_\i - \chi_\j
\end{equation}
to get the action:
\begin{eqnarray}\label{TriSineGordon2}
S & = & \frac{e^2}{2}\sum_{\bond{\i\j}} A_{\i\j}^2
        + \frac{g}{2}\sum_{\textrm{plaq.}}\bigl(\textrm{curl}A\bigr)^2 \nonumber \\
  & - & K \sum_{\bond{\i\j}} \cos 2\pi\Bigl( \chi_\i - \chi_\j + A_{\i\j} +
                \frac{1}{2}A^{(0)}_{\i\j} \Bigr) \nonumber \\
  & - & \gamma \sum_\i \cos 2\pi\bigl( \chi_\i - \chi^{(0)}_\i \bigr) \ .
\end{eqnarray}
Note that the term proportional to $e^2$ appears as a ``mass'' term for the gauge field $A_{\i\j}$. We therefore integrate out $A_{\i\j}$. This may be explicitly done in the following manner. We formally expand $\exp(-S)$ in powers of $K$, and decompose every cosine factor from the expansion using the Euler's formula $2\cos\theta = \exp(i\theta) + \exp(-i\theta)$. The expansion takes the following form:
\begin{eqnarray}
e^{-S} & = & \exp\Bigl\lbrace \gamma \sum_\i \cos 2\pi\bigl( \chi_\i - \chi^{(0)}_\i 
             \bigr) \Bigr\rbrace \times \nonumber \\
       & \times & \sum_{\lbrace m_{\i\j}\rbrace} C_{\lbrace m_{\i\j}\rbrace}
             \exp\Bigl\lbrace -\Bigl[ \frac{e^2}{2}\sum_{\bond{\i\j}} A_{\i\j}^2 \\
    & & + \frac{g}{2}\sum_{\textrm{plaq.}}\bigl(\textrm{curl}A\bigr)^2 \nonumber \\
    & & + \sum_\i 2\pi i m_{\i\j} \Bigl( 
          \chi_\i - \chi_\j + A_{\i\j} + \frac{1}{2}A^{(0)}_{\i\j} \Bigr)
          \Bigr] \Bigr\rbrace \nonumber \ ,
\end{eqnarray}
where $m_{\i\j}$ are integers, and $\lbrace m_{\i\j} \rbrace$ denotes their distribution on the whole lattice. This is a Gaussian form, and $A_{\i\j}$ can be easily integrated out. Clearly, this causes $\lbrace m_{\i\j} \rbrace$ dependent renormalization of the $C_{\lbrace m_{\i\j} \rbrace}$ factors, which in turn corresponds to renormalization of $K$. After resummation over $m_{\i\j}$, an XY model is obtained at the lowest order in (renormalized) $K$:
\begin{eqnarray}\label{TriXY}
S & = & - K \sum_{\bond{\i\j}} \cos 2\pi\Bigl( 
          \chi_\i - \chi_\j + \frac{1}{2}A^{(0)}_{\i\j} \Bigr) \\
  & &  - \gamma  \sum_\i \cos 2\pi\bigl( \chi_\i - \chi^{(0)}_\i \bigr)
       - \mathcal{O}\bigl( K^2, \gamma^2 \bigr) \ . \nonumber
\end{eqnarray}
The variables $2\pi\chi_\i$ should be treated as angles. Adding integers to $\chi_\i$ in the gauge theory ~(\ref{TriAction2}) can always be compensated by a gauge transformation, so that the gauge inequivalent states in ~(\ref{TriXY}) correspond to different fractional parts of $\chi_\i$. This is why the obtained effective theory is an XY model. Physically, the dual field $2\pi\chi_\i$ represents the phase of an operator that creates $2\pi$ units of vorticity in the bosonic matter field of the original honeycomb lattice U(1) gauge theory ~(\ref{HonU1}), and thus must fundamentally be a phase, after the redundant gauge degrees of freedom have been removed. The structure of this theory is such that the XY field fluctuates in presence of a fixed staggered flux, and there is also an external field that apparently explicitly breaks the XY (and apparently also lattice) symmetries (see Fig.~\ref{XYExtField}). The physical meaning of this explicit symmetry breaking term is as follows. In the original honeycomb lattice gauge theory $2\pi$ vortices of the bosonic matter field carry $2\pi$ units of the U(1) gauge flux. As the theory is {\em compact}, instanton events where this gauge flux changes by $2\pi$ are allowed - thus the vorticity is not strictly conserved. The explicit XY symmetry breaking term in the dual representation precisely describes these instanton events. 

What is the actual symmetry of this action? As discussed above, when $\gamma = 0$ there is a global XY symmetry that is apparently broken when $\gamma \neq 0$. However, a discrete $\mathbb{Z}_3$ subgroup of the global XY symmetry survives when combined with translation by one lattice spacing. For instance, the action is invariant under translation along the horizontal $x$-axis by one unit:
\begin{equation}\label{LatticeTranslation}
\chi_r \rightarrow \chi_{r - \hat{x}} + \frac{2}{3} \ .
\end{equation}
Similar transformation properties obtain under other translations (as well as the other lattice symmetries).

\subsection{Phase Diagram}

Armed with the dual formulation, we can now see qualitatively why a translation invariant phase is possible in this model. Consider the limit $\gamma = 0$. Then the resulting global XY model will (for $K$ small enough) possess a disordered phase with short-ranged correlations for the $\chi_\i$ field. Upon increasing $K$ a transition to an ordered phase with some pattern of XY ordering will occur. Now consider turning on a small $\gamma$. Its effects will be innocuous in the small-$K$ disordered phase. In particular, the discrete $\mathbb{Z}_3$ symmetry of $\frac{2}{3}$ shifts of the $\chi_\i$ field, which realize lattice translations ~(\ref{LatticeTranslation}), will continue to stay unbroken. This is therefore a translation-invariant phase. It is readily seen that this phase is also invariant under all other lattice symmetries. The $\gamma$ term will have a much more significant effect in the large-$K$ ordered phase, where the XY symmetry is completely broken: it will pin the overall orientation of the XY ordering pattern. Lattice symmetries are consequently expected to be broken in such a phase. 

The existence of the disordered translation-invariant phase is our primary conclusion. How do we think about it in terms of the original gauge theory? From the physical discussion above, the field $e^{2i\pi\chi_\i}$ creates $2\pi$ flux of the original gauge field, which in turn is bound to a $2\pi$ vortex in the phase of the bosonic matter field. Thus the disordered phase is to be thought of as a ``Higgs'' phase where the bosonic matter fields have condensed - which gaps out their vortices. Indeed, our gauge model is closely related to a similar one discussed in Ref. \cite{SJ} (the ``$N =1$ SJ model'') where similar phenomena were shown to arise. Note that the Higgs phase is preferred by the boson hopping term in the gauge theory. Thus it is reasonable to expect that this phase is realized in the limit $\Gamma \gg t$. This expectation is indeed consistent with the numerical results of Ref. \cite{FrIsing1,FrIsing2}. 

\begin{figure}
\subfigure[{}]{\includegraphics[height=1.2in]{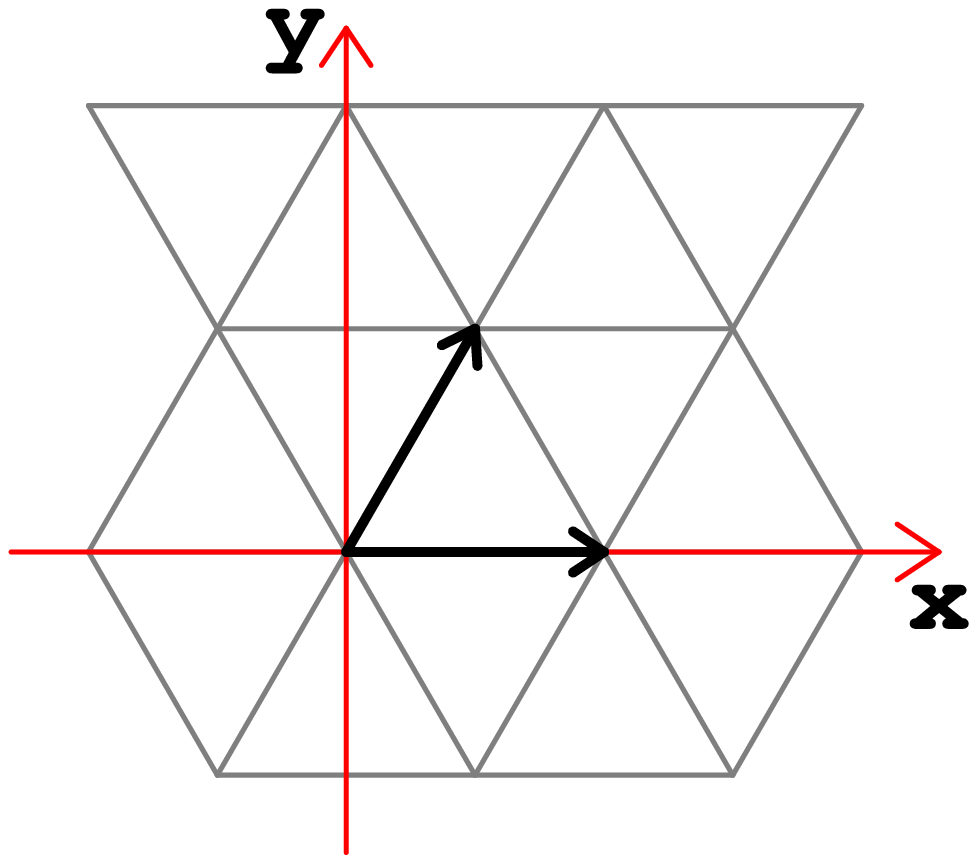}}
\subfigure[{}]{\includegraphics[height=1.2in]{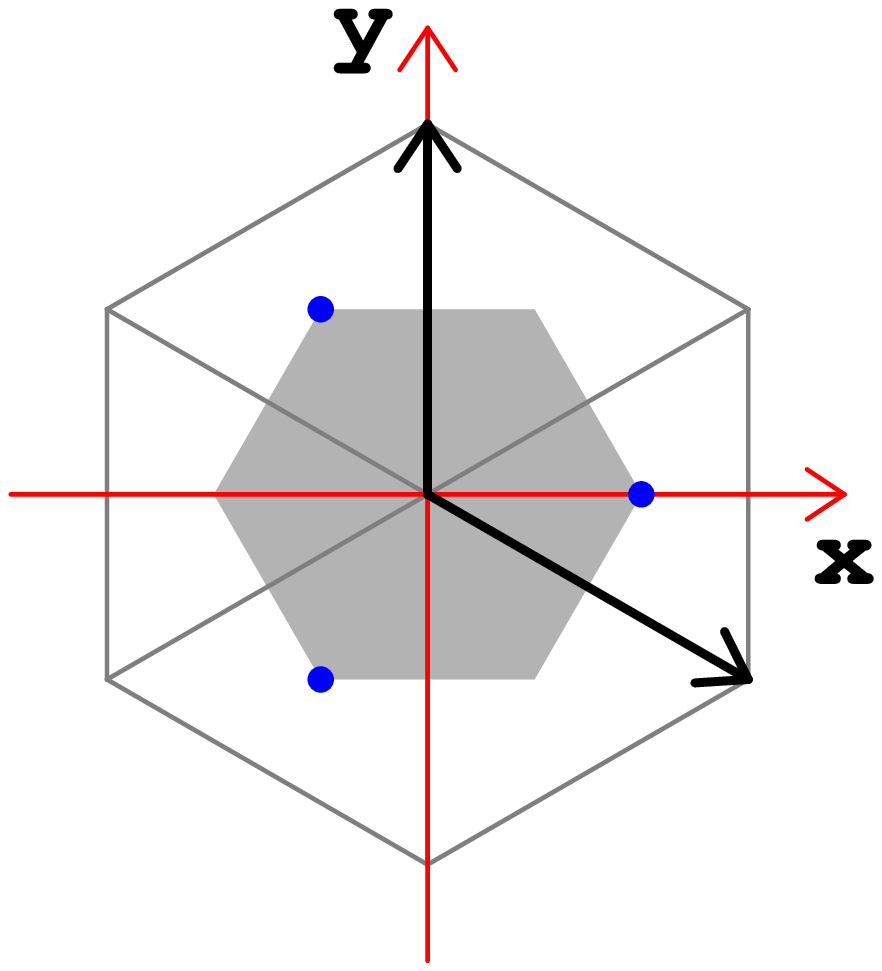}}
\caption{\label{TriBZ}(a) Direct triangular lattice, and a choice of primitive vectors: $\boldsymbol{a_1} = \uv{x}, \boldsymbol{a_2} = \frac{1}{2}\uv{x} + \frac{\sqrt{3}}{2} \uv{y}$. (b) Reciprocal lattice with primitive vectors $\boldsymbol{b_1} = 2\pi(\uv{x} - \frac{1}{\sqrt{3}} \uv{y}), \boldsymbol{b_2} = 2\pi \cdot \frac{2}{\sqrt{3}} \uv{y}$. Shaded region represents the first Brillouin zone. Emphasized corners of the Brillouin zone represent the wavevector $\boldsymbol{Q} = \frac{4\pi}{3} \uv{x}$ (three equivalent points).}
\end{figure}

We can formally back up this discussion by considering an alternate soft-spin version of the dual XY model. To that end we define:
\begin{equation}\label{Phi}
\Phi_\rv \sim e^{2\pi i \chi_\rv} \ .
\end{equation}
We have labelled the triangular lattice sites by their corresponding vectors $\rv$. After substituting this into the XY model ~(\ref{TriXY}) and relaxing the ``hard-spin'' condition on $\Phi_{\rv}$ we obtain a ``soft-spin'' lattice theory:
\begin{eqnarray}\label{SoftSpinTheory}
S & = & -\frac{K}{2} \sum_{\bond{\rv\rvp}}
    \bigl( \Phi^*_\rv e^{-i\pi A^{(0)}_{\rv\rvp}} \Phi^{\phantom{*}}_\rvp + h.c. \bigr)
    \nonumber \\
  & & + r|\Phi_{\rv}|^2 + u|\Phi_{\rv}|^4 + \cdots \\
  & & - \gamma \sum_\rv \bigl( e^{-i\Qv\rv} \Phi^{\phantom{*}}_\rv + h.c. \bigr) 
      - \cdots \ . \nonumber
\end{eqnarray}
The quadratic part of this action can be diagonalized to obtain:
\begin{equation}\label{QuadPart}
\sum_\qv \left(-K \varepsilon_\qv +r \right) |\Phi_\qv|^2 \ ,
\end{equation}
where
\begin{eqnarray}\label{XYDisp}
\varepsilon_\qv & = & \cos \Bigl( q_x - \frac{\pi}{3} \Bigr) +
  \cos \Bigl( \frac{-q_x + \sqrt{3} q_y}{2} - \frac{\pi}{3} \Bigr) \nonumber \\
  & + & \cos \Bigl( \frac{-q_x - \sqrt{3} q_y}{2} - \frac{\pi}{3} \Bigr) \ .
\end{eqnarray}
It is straight-forward to show that $\varepsilon_\qv$ is maximized at two differend wavevectors: $\qv=0$, and $\qv=-\Qv$, where $\Qv$ is the wavevector that describes spatial variation of the external XY field:
\begin{equation}\label{QQ}
\Qv=\frac{4\pi}{3}\uv{x} \quad , \quad e^{2\pi i \chi^{(0)}_\rv}=e^{i \Qv \rv} \ .
\end{equation}
In order to aid understanding of this special wavevector, we plot its location on the reciprocal lattice in the Fig.~\ref{TriBZ}.

In the following, we focus on the limit $K \gg \gamma$. The fluctuations at low energies will be dominated by the modes in vicinity of the two different wavevectors, $\qv = 0$ and $\qv = -\Qv$. Such fluctuations can be expressed by:
\begin{equation}\label{Phi2}
\Phi_\rv = \psi_{1,\rv} + e^{-i\Qv\rv} \psi_{2,\rv} \ ,
\end{equation}
where $\psi_{1,\rv}$ and $\psi_{2,\rv}$ are fields that vary slowly on the scale of the lattice spacing. It is then useful to go to a continuum limit which focuses on the long wavelength fluctuations of these two fields. This continuum theory must be expressed in terms of $\psi_{1,\rv}$ and $\psi_{2,\rv}$ only, and must be invariant under all symmetry transformations of the lattice theory ~(\ref{SoftSpinTheory}). First, we note that the quadratic part ~(\ref{QuadPart}) is invariant under global XY rotations $\Phi_\rv \to e^{i\theta} \Phi_\rv$, as well as all lattice symmetry transformations at small deviations $\kv$ from the wavevectors $\qv \in \lbrace 0, -\Qv \rbrace$:
\begin{equation}
(\forall \alpha) \qquad -K\varepsilon_{\qv+\kv} =
   \textrm{const.} + \kappa k^2 + \mathcal{O}(k^3) \ .
\end{equation}
In addition, there is also a symmetry under exchanging the two kinds of slowly varying fields, $\psi_{1,\rv} \leftrightarrow \psi_{2,\rv}$. On the other hand, as discussed in the previous section, the $\gamma$ term (and various similar higher order terms) break the XY symmetry explicitly, and reduce the symmetry group to a discrete set of transformations. It is easy to show that under a unit translation by $\rvo$ the fields transform as:
\begin{equation}\label{Transf1}
\psi_{1,\rv} \to e^{i\Qv\rvo} \psi_{1,\rv-\rvo} \quad , \quad
\psi_{2,\rv} \to e^{-i\Qv\rvo} \psi_{2,\rv-\rvo} \ .
\end{equation}
Note that this rotates the phases of $\psi_1$ and $\psi_2$ by $120$ degrees in opposite directions. Similarly, the symmetry transformation $\psi_{1,\rv} \leftrightarrow \psi_{2,\rv}$ of the quadratic part of the action must be accompanied by lattice inversion $\rv \to -\rv$ in order to keep the whole action ~(\ref{SoftSpinTheory}) invariant:
\begin{equation}\label{Transf2}
\psi_{1,\rv} \to \psi_{2,-\rv} \quad , \quad \psi_{2,\rv} \to \psi_{1,-\rv} \ .
\end{equation}
The remaining lattice transformations do not give rise to further reduction of the symmetry group. We may now write down a Landau-Ginzburg effective field theory, symmetric under transformations ~(\ref{Transf1}) and ~(\ref{Transf2}):
\begin{eqnarray}\label{FieldTheory}
S & = & \int\dd\kv \Bigl[ ( r + \kappa k^2 ) \Bigl(
    |\psi^{\phantom{*}}_{1,\kv}|^2 + |\psi^{\phantom{*}}_{2,\kv}|^2 \Bigr) + \nonumber \\
  & & ( a + b k^2 ) \Bigl(
        \psi^{\phantom{*}}_{1,\kv} \psi^{\phantom{*}}_{2,-\kv} + h.c. \Bigr) \Bigr] \\
  & - & \int\dd\rv \Bigl[ \beta \Bigl(
        \psi^2_{1,\rv} \psi^{\phantom{*}}_{2,\rv} + \psi^2_{2,\rv} \psi^{\phantom{*}}_{1,\rv}
      + h.c. \Bigr) + \nonumber \\
  & & \gamma \Bigl(
        \psi^3_{1,\rv} + \psi^3_{2,\rv} + h.c. \Bigr) \Bigr] \nonumber \\
  & + & \int\dd\rv \Bigl[ u \Bigl(
        |\psi^{\phantom{*}}_{1,\rv}|^4 + |\psi^{\phantom{*}}_{2,\rv}|^4 \Bigr)
      + u' |\psi^{\phantom{*}}_{1,\rv}|^2 |\psi^{\phantom{*}}_{2,\rv}|^2 +
        \nonumber \\ 
  & & u'' \Bigl( \psi^{\phantom{*}}_{1,\rv} \psi^{\phantom{*}}_{2,\rv} + h.c.
        \Bigr) \Bigl(
        |\psi^{\phantom{*}}_{1,\rv}|^2
      + |\psi^{\phantom{*}}_{2,\rv}|^2 \Bigr)
      \Bigr] + \cdots \ . \nonumber
\end{eqnarray}
It is clear that at least for the coupling $r$ sufficiently large and positive a stable disordered phase will exist where the $\mathbb{Z}_3$ symmetry is unbroken. This then corresponds to the translation symmetric phase of the original lattice model.

Ordered phases are also possible in this theory. In order to reveal their structure, we simply consider the static classical XY field configurations that minimize the action of the effective XY model ~(\ref{TriXY}). Let us ignore the terms $\mathcal{O} (K^2,\gamma^2)$, and restrict our attention to configurations where the value $2\pi\chi_{\rv}$ depends only on the sublattice to which the site $\rv$ belongs. The energy per unit-cell of such configurations is:
\begin{eqnarray}\label{XYSPEq}
E & = & -3K \Bigl[
        \cos 2\pi \Bigl(\chi_1 - \chi_2 + \frac{1}{6} \Bigr) \\
  & & + \cos 2\pi \Bigl(\chi_2 - \chi_3 + \frac{1}{6} \Bigr) \nonumber \\
  & & + \cos 2\pi \Bigl(\chi_3 - \chi_1 + \frac{1}{6} \Bigr) \Bigr] \nonumber \\
  & - & \gamma \sum_{n=1}^3 \cos 2\pi \bigl(\chi_n - \chi^{(0)}_n \bigr) \ , \nonumber
\end{eqnarray}
where we have labeled the sublattices by $1, 2, 3$ in such a way that the background gauge field vector $A^{(0)}_{\rv\rvp}$ circulates in the direction $1 \to 2 \to 3 \to 1$. Two kinds of states can be found. For sufficiently large $\gamma$, the XY field simply follows the ``external'' XY field: $\chi_{\rv} = \chi^{(0)}_{\rv}$, and an XY ``spin density wave'' is established at the wavevector $\Qv$. Naively, the XY rotation and lattice symmetries appear explicitly broken in this state. However, the angular order parameter $2\pi\chi_{\rv} = 2\pi\chi^{(0)}_{\rv}$ is invariant under the lattice translations ~(\ref{LatticeTranslation}) (and other lattice transformations), so that this state of the effective XY model corresponds to the disordered state of the original spin model. On the other hand, if $\gamma=0$ the ordering will be determined by the nearest-neighbor interaction $K$. As we have argued before, there are two ordering wavevectors ($\qv=0$ and $\qv=-\Qv$) preferred by the $K$ term, but neither of them coincides with the wavevector that describes spatial variations of the external XY field $\chi^{(0)}_{\rv}$. Consequently, any small non-zero $\gamma$ will introduce frustration, and deform the spontaneously ordered XY ``spin density wave'' preferred by the $K$ term. Even though the precise description of the ordering pattern is complicated, it is at least apparent that the XY fields $\chi_{\rv}$ align with the external field $\chi^{(0)}_{\rv}$ on one sublattice of the triangular lattice, while on the other two sublattices they cant toward the external field, simultaneously trying to preserve the preferred ordering wavevector. Arbitrary choice of sublattice for the alignment gives rise to a three-fold degeneracy, so that the $\gamma$ term breaks the continuous XY symmetry down to a $\mathbb{Z}_3$ subgroup, associated with lattice transformations (as evident in the Equation ~(\ref{Transf1})). However, choice of the ``parent'' ordering wavevector ($\qv=0$ or $\qv=-\Qv$) for sufficiently small $\gamma$ is also available, and we will briefly discuss its physical origin.

One symmetry transformation that we have ignored so far is the global spin-flip in the original spin model on the Kagome lattice: $S^z_\ik \to -S^z_\ik$. It is straight forward to trace back how this transformation affects the quantities of the U(1) gauge theory, and its dual theory on the triangular lattice. For example, at the level of Equation ~(\ref{TriAction}), which describes the dual theory on the triangular lattice with integer-valued gauge field $A_{\i\j}$, and integer-valued $\chi_\i - \chi^{(0)}_\i$, the global spin-flip corresponds to:
\begin{eqnarray}\label{GlobalSpinFlip}
A_{\i\j} & \longrightarrow & \bigl( A^{(0)}_{\i\j} + \chi^{(0)}_\j - \chi^{(0)}_\i
   \bigr) - A_{\i\j} \nonumber \\
\chi_\i & \longrightarrow & -\chi_\i - \chi^{(0)}_\i \ .
\end{eqnarray}
Note that the action ~(\ref{TriAction}) is invariant under this transformation, and that the integer constraints are not affected. Only the second part of this transformation (involving the $\chi_\i$ field) survives in the effective XY theory ~(\ref{TriXY}). If we now turn to the continuum limit, we find from ~(\ref{QQ}), ~(\ref{Phi2}) and ~(\ref{GlobalSpinFlip}) that the global spin-flip is represented by:
\begin{equation}\label{Transf3}
\psi^{\phantom{*}}_{1,\rv} \longrightarrow \psi^*_{2,\rv} \quad , \quad
\psi^{\phantom{*}}_{2,\rv} \longrightarrow \psi^*_{1,\rv} \ .
\end{equation}
Since the fields $\psi_{1,\rv}$ and $\psi_{2,\rv}$ describe the XY ``spin density waves'' at wavevectors $\qv=0$ and $\qv=-\Qv$ respectively, we see that the spin-flip formally exchanges these two kinds of order in the effective theory. Apparently, this transformation is completely independent from lattice translations. Therefore, the choice of the ``parent'' ordering wavevector ($\qv=0$ or $\qv=-\Qv$) in the spontaneously ordered phase must correspond to the choice of direction of the global magnetization. 

\begin{figure}
\oldnewfig
{\includegraphics[height=2.2in]{phase-diagram.eps}}
{\includegraphics[height=2.2in, viewport=270 520 565 720, clip]{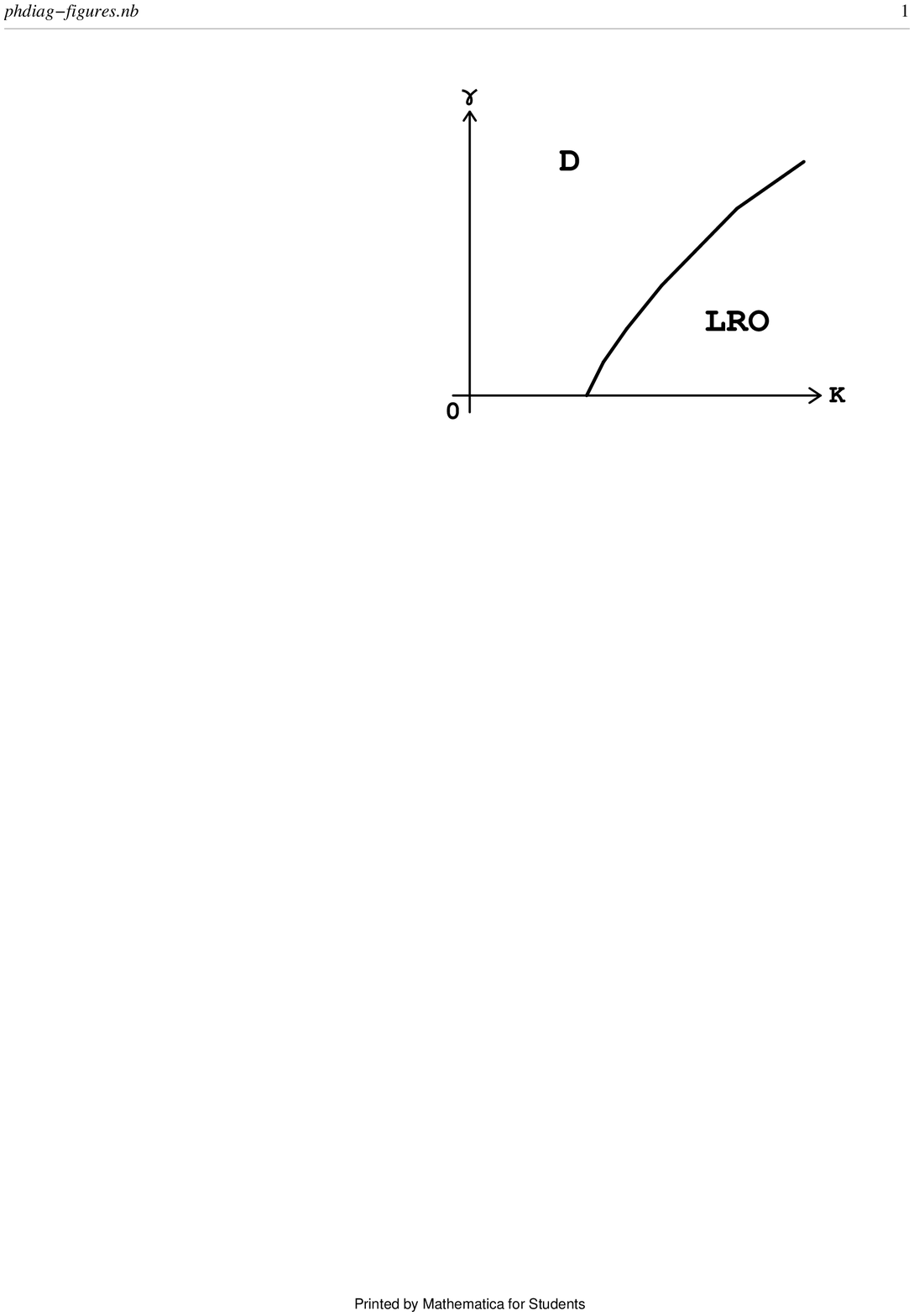}}
\caption{\label{XYPD}Schematic phase diagram of the effective XY model ~(\ref{TriXY}). D corresponds to the disordered phase of the original Kagome spin model. LRO is a spontaneously long-range ordered and magnetized phase.}
\end{figure}

\begin{figure}
\includegraphics[height=1.7in]{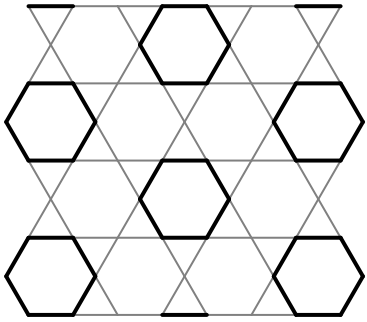}
\caption{\label{IsingKagVBC}The most symmetric pattern of three-fold translational symmetry breaking on the Kagome lattice. Roughly speaking, the six spins on the emphasized hexagons are alternating, and coherently resonating as a singlet $\frac{|\uparrow \downarrow \uparrow \downarrow \uparrow \downarrow \ra + |\downarrow \uparrow \downarrow \uparrow \downarrow \uparrow \ra}{\sqrt{2}}$, while the remaining spins are ferromagnetically aligned with each another and break the global spin-flip symmetry. Note that in the Hamiltonian ~(\ref{HonEff}) the energy is reduced by $\mathcal{O}(t)$ on every resonating hexagon.}
\end{figure}

We can finally sketch the phase diagram of the effective XY model ~(\ref{TriXY}), as shown in the Fig.~\ref{XYPD}. The phase D is invariant under lattice translations ~(\ref{LatticeTranslation}) and global spin-flip ~(\ref{GlobalSpinFlip}), so that it corresponds to the disordered phase of the original spin model ~(\ref{HonEff}), realized for $\Gamma \gg t$. The spontaneously long-range ordered phase LRO is obtained in the opposite limit of strong hexagon ring-exchange $\Gamma \ll t$. We argued that this phase has global Ising magnetization, while the translation symmetry is broken in a three-fold degenerate manner. This naturally suggests a microscopic description of the order parameter given in the Fig.~\ref{IsingKagVBC}. We would like to note that such a state exactly corresponds to the state with one-third of saturated magnetization on the Kagome lattice that was found responsible for plateaus in the magnetization curves of some Kagome-based systems. \cite{MagnPlateau} The phase transition is most likely of the second order (if we had considered a small \emph{longitudinal} field in the original Kagome lattice spin model, the ultimate continuum limit would have been an XY model with a three-fold XY anisotropy, which has a second order transition). 

\section{Discussion}

The theory of the Kagome lattice quantum Ising model that we have presented in this paper reveals that a disordered ground state is found for weak transverse field. This disordered phase breaks no symmetries and is not topologically ordered either. It is therefore expected to be smoothly connected to the completely uncorrelated phase at large transverse field. The same conclusion has been strongly suggested by Monte Carlo simulations, \cite{FrIsing1, FrIsing2} where no phase transition was detected as the strength of transverse field was varied.

Such disordered phases seem to be exceptions rather than the rule for a large set of two-dimensional frustrated lattice models with similar structure. If the classical ground state manifold is macroscopically degenerate in a system with discrete degrees of freedom, a typical situation is that quantum fluctuations reduce the degeneracy down to the one associated with broken lattice symmetries, creating ``order-by-disorder''. Examples include the hard-core quantum dimer and Ising models on most studied simple lattices, \cite{FrIsing1, FrIsing2}. The analysis in this paper provides a route to understanding the differences between these common situations with order-by-disorder and the Kagome quantum Ising antiferromagnet. We have demonstrated how the Kagome lattice quantum Ising model can be described by a compact U(1) gauge theory with some fixed background charge at each lattice site. Such a gauge theoretic description has been useful in studies of similar Ising models on other frustrated lattices. The crucial distinction between the Kagome and other common systems is in the fact that the U(1) gauge theory of the Kagome system contains a dynamical matter field. Without a matter field, these gauge theories of the frustrated magnets in two-dimensions always ultimately live in a confined phase that breaks translation symmetry. The latter is caused by the fixed background charge in the gauge theory. However, in the presence of dynamical matter fields a translation invariant (``Higgs'') phase is generally possible. In situations where the matter field has gauge charge $2$ (as happens in the gauge theoretic description of quantum dimer models on non-bipartite lattices) such a Higgs phase also possesses topological order and associated ``vison'' excitations: for instance, there is a non-trivial ground state degeneracy on topologically non-trivial manifolds. In the problem discussed in this paper the matter field had gauge charge $1$.  The resulting Higgs phase, while translation invariant, is topologically trivial. This then lends strong support to the conjecture in Ref.\cite{FrIsing1} that weak transverse fields on the Kagome Ising magnet lead to a phase that is smoothly connected to the trivial paramagnet that obtains at strong transverse fields. 

In fact, topological triviality of the discovered disordered phase is a very interesting detail. From the beginning of the quest for interesting quantum spin liquids, the frustrated magnets had been looked upon with great hope. In particular, the systems with extremely large classical ground state degeneracy have attracted considerable attention. The pyrochlore and Kagome lattices are thought to be ideal candidates for the spin liquid, because their corner-sharing structure provides such extremely large degeneracy, which is then dramatically lifted even by weak quantum fluctuations. However, in the case of the Kagome lattice transverse-field quantum Ising model, the corner-sharing geometry seemingly makes the lattice virtually disconnected. Even though there is no long-range order in the ground state, the obtained disordered phase is not a topologically ordered spin liquid. One may then speculate that a frustrated, but somewhat more connected lattice is probably a better platform to seek topologically non-trivial spin liquids. Another question is if a more correlated spin dynamics, such as the one found in anisotropic easy-axis Heisenberg models, is more likely to yield such interesting spin liquids. These issues are left open for future work.  

This research was supported by the National Science Foundation under the grant DMR-0308945. We would also like to acknowledges funding from the NEC Corporation, the Alfred P. Sloan Foundation (T.S) and an award from The Research Corporation (T.S.).

% -----------------------------------------------
% |                REFERENCES                   |
% -----------------------------------------------

\end{document}